\documentclass[10pt]{article}
\title{Frequency and Wavelength of Light\\in Relativistically Rotating Frames}
\author{Robert D. Klauber\\1100 University Manor Dr., 38B, Fairfield, IA 52556, USA\\email: rklauber@netscape.net}
\date{July 23, 2002}
\usepackage {graphicx}
\usepackage {longtable}
\usepackage{hyperref}
\begin{document}
\maketitle
\begin{abstract}


Non-time-orthogonal frame analysis is applied to determine the frequency and 
wavelength of light as observed i) in a relativistically rotating frame when 
emission is from a source fixed in the non-rotating frame, ii) in a 
non-rotating frame when emission is from a source fixed in the rotating 
frame, and iii) when both source and observer are fixed in the rotating 
frame and the source emission direction varies with respect to the rotating 
frame. Appropriate Doppler effects are demonstrated, and second order 
differences from translating (time-orthogonal) frame analysis are noted.

04.20.-q, 04.20.Cv 
\end{abstract}

\bigskip

\section{INTRODUCTION}

\subsection{Background}

An analysis\cite{Robert:1998} \cite{Robert:1999} \cite{Robert:1} has been 
carried out of the non-time-orthogonal metric (i.e., $g_{0i} \ne 0$, time is 
not orthogonal to space) obtained when one makes a straightforward 
transformation from the lab to a relativistically rotating frame. Rather 
than assuming, as have other researchers, that it is then necessary to 
transform to locally time orthogonal (i.e., time is orthogonal to space) 
frames, one can proceed by considering the non-time-orthogonal (NTO) metric 
to be a physically valid representation of the rotating frame.

When this is done, one finds the usual time dilation and mass-energy 
dependence\cite{Ref:1} on tangential speed $\omega r$, in full accord with 
the test data from numerous cyclotron experiments. One also finds 
resolutions of paradoxes inherent in the traditional analytical treatment of 
rotating frames. Further, the analysis predicts at least one experimental 
result\cite{Ref:2} \cite{Ref:3} that, in the context of the traditional 
analysis, has heretofore been considered inexplicable.

NTO frame analysis is in full accord with fundamental principles of 
relativity theory, and makes many of the same predictions as the traditional 
analysis for rotating frames. It does not conflict with recognized analyses 
of time-orthogonal (TO) frames, including those described by Lorentz, 
Schwarzchild, and Friedman metrics. Just as for TO frames, the NTO line 
element remains invariant, and differential geometry reigns as the 
appropriate descriptor of non-inertial systems.

However, NTO analysis does predict some behavior that may seem strange from 
a traditional relativistic standpoint, though it appears corroborated by 
both gedanken and physical experiments\cite{Neil:1997} \cite{Neil:2002} 
\cite{Robert:2}. In particular, NTO analysis finds the specific result for 
the speed of light in the circumferential direction for rotating (NTO) 
frames to be non-invariant, non-isotropic, and equal to\cite{Ref:4}

\begin{equation}
\label{eq1}
u_{light,circum} \,\; = \;\,\,\frac{c\pm \omega r}{\sqrt {1 - (\omega r)^2 / 
c^2} } = \frac{c\pm v}{\sqrt {1 - v^2 / c^2} },
\end{equation}

\noindent
where the sign before $v=\omega r$ depends on the circumferential direction of 
the light ray at $r$ relative to the tangential speed $v$. Note the 
circumferential light speed in the rotating frame varies to first order with 
$\omega r$.

This result agrees with Ashby's research with the global positioning system. 
Ashby notes `` .. the principle of the constancy of $c$ cannot be applied in a 
rotating reference frame ..''\cite{Neil:2002}. He also states ``Now 
consider a process in which observers in the rotating frame attempt to use 
Einstein synchronization [constancy of the speed of light] ..... Simple 
minded use of Einstein synchronization in the rotating frame ... thus leads 
to a significant error''\cite{Neil:1997}. 

\subsection{Overview}
\label{subsec:overviewult}

In the present article NTO analysis is used to determine frequency and 
wavelength of light i) emitted from a source in the lab and observed from 
the rotating frame, ii) emitted from a source in the rotating frame and 
observed in the lab, and iii) emitted from a source in the rotating frame 
that turns relative to the rotating frame. Consistency is evidenced in that 
multiplication of frequency and wavelength thus obtained for the rotating 
frame yields (\ref{eq1}).

As background, and as an aide for comparison, Section 
\ref{sec:newtonian} provides a summary of speed, frequency, 
wavelength, and Doppler shift of waves propagating through elastic media for 
various observers in a Newtonian universe. Section 
\ref{sec:waves} provides a similar summary for waves 
propagating in vacuum (i.e., non-elastic waves) for both Newtonian and 
Lorentzian observers. The transformation between the lab and the rotating 
frames, as well as the resulting NTO metric for the rotating frame, as 
derived in cited references, are listed in Section 
\ref{sec:rotating}. That section also includes a summary of the 
differences between the generalized coordinate components of vectors used in 
mathematical analysis and the physical components that equal the values 
actually measured in physical experiments. Section 
\ref{sec:rotating} then prescribes the method for converting 
coordinate components to physical components and vice versa, as well as the 
procedure for using such conversions to solve problems of a most general 
nature. The mathematical relations and methodology of Section 
\ref{sec:rotating} are subsequently applied in Section 
\ref{sec:mylabel2} to determine appropriate wave frequencies and 
wavelengths of light as seen by lab and rotating frame observers. Section 
\ref{sec:turning} addresses the case where both source and observer 
are in the rotating frame and the light source is turned relative to that 
frame.

\section{NEWTONIAN WAVES IN ELASTIC MEDIA}
\label{sec:newtonian}

With minimal comment we present Table 1, a rather elementary 
summary\cite{David:1974} of elastic wave propagation in Galilean frames, 
which will prove of value for comparison with the results of Section 
\ref{sec:mylabel2}. Prior to Section \ref{sec:turning}, we 
treat only cases in which motions of the observer, the source, and the 
medium, if any, are along the line of sight between the observer and source. 
In Table 1 the light source is to the left of the observer, the wave travels 
toward the right, $v_{m}$ is the speed of the wave within the elastic medium 
(i.e., relative to the medium), and $v$ is the speed of the source toward 
(approaching) the observer. Positive displacement, and hence velocity, is to 
the right. Quantities in the source frame $K$ are unprimed, in the observer 
frame ${K}'$ are primed, and in the medium have a subscript $m$. In all tables 
presented herein source and observer receding from one another implies $v$ 
becomes $ - v$ in all blocks within a given table.

Note that in all cases multiplication of frequency by wavelength as seen 
either in the source frame, or in the observer frame, results in the correct 
wave speed for the given frame. Note further in Table 1 the second order 
difference in Doppler effect seen by the observer when the observer is fixed 
in the medium (Case 2) as opposed to when the source is fixed in the medium 
(Case 1).

\begin{center}
\textbf{Table 1. Summary: Waves in Elastic Media in Galilean Frames}
\end{center}

\begin{longtable}[htbp]
{|p{85pt}|p{112pt}|p{157pt}|p{130pt}|}
a & a & a & a  \kill
\hline
& 
\textbf{Case 1}& 
\textbf{Case 2}& 
\textbf{Case 3} \\
\hline
Source frame $K$& 
Fixed to medium& 
Not fixed to medium& 
Not fixed to medium \\
\hline
Observer ${K}'$& 
Not fixed to medium & 
Fixed to medium& 
Not fixed to medium  \\
\hline
Motion& 
Source (medium) \par toward observer at $v$& 
Source toward \par observer (medium) at $v$& 
Source, observer both at rest. Medium at $v$. \\
\hline
Wave speed& 
$V=v_{m}$ \par ${V}' = v_m + v$& 
$V=v_{m}$ -- $v$ \par ${V}' = v_m $& 
$V = {V}' = v_m + v$ \\
\hline
Wave length& 
${\lambda }' = \lambda $& 
${\lambda }' = \lambda $& 
${\lambda }' = \lambda = \lambda _m $ \\
\hline
Frequency& 
${f}' = f\left( {1 + v / v_m } \right)$& 
${f}' = \frac{f}{1 - v / v_m }$& 
${f}' = f = f_m \left( {1 + v / v_m } \right)$ \\
\hline
Doppler shift \par observer sees& 
${f}' = f\left( {1 + v / v_m } \right)$& 
${f}' = f\left( {1 + v / v_m + v^2 / v_m^2 + ...} \right)$& 
None \\
\hline
\label{tab1}
\end{longtable}

\section{WAVES WITHOUT ELASTIC MEDIA}
\label{sec:waves}

Table 2 summarizes\cite{Ibid:1} the behavior of light waves that propagate 
through vacuum without an underlying supporting medium for both a Newtonian 
and a relativistic universe. Note that for Lorentz frames the wavelength 
appears to the observer to have different length than it does in the source 
frame. As for elastic waves, in each case multiplication of frequency times 
wavelength equals wave speed for a given frame. The second order dependence 
for the relativistic Doppler shift and how it differs from either Case 1 or 
Case 2 in Table 1 is well known, and thereby provides a means to test 
special relativity\cite{Herbert:1941}.

\bigskip

\begin{center}
\textbf{Table 2. Summary: Waves without Media, Galilean and Lorentzian}
\end{center}

\begin{longtable}[htbp]
{|p{104pt}|p{162pt}|p{162pt}|}
a & a & a  \kill
\hline
& 
\textbf{Galilean Frames}& 
\textbf{Lorentz Frames} \\
\hline
Source K& 
No medium& 
No medium \\
\hline
Observer K$\prime $& 
No medium & 
No medium  \\
\hline
Motion& 
Source toward observer at $v$& 
Source toward observer at $v$ \\
\hline
Wave speed& 
$V = c$ \par ${V}' = c + v$& 
$V = c$ \par ${V}' = c$ \\
\hline
Wave length& 
${\lambda }' = \lambda $& 
${\lambda }' = \lambda \frac{\sqrt {1 - v / c} }{\sqrt {1 + v / c} } = \lambda \frac{1 - v / c}{\sqrt {1 - v^2 / c^2} }$ \\
\hline
Frequency& 
${f}' = f\left( {1 + v / c} \right)$& 
${f}' = f\frac{\sqrt {1 + v / c} }{\sqrt {1 - v / c} } = f\frac{1 + v / c}{\sqrt {1 - v^2 / c^2} }$ \\
\hline
Doppler shift \par observer sees& 
${f}' = f\left( {1 + v / c} \right)$& 
${f}' = f\left( {1 + v / c + \frac{1}{2}v^2 / c^2 + ...} \right)$ \\
\hline
\label{tab2}
\end{longtable}

\section{ROTATING FRAMES}
\label{sec:rotating}

\subsection{Transformation and metric}

We adopt notation in which the Minkowski metric for a Lorentz frame has form 
\textit{$\eta $}$_{\alpha \beta }$ = diag (-1,1,1,1). For the rotating frame analysis we 
employ cylindrical coordinates with (\textit{cT,R,$\Phi $,Z) }for the lab frame $K$ and (\textit{ct,r,}$\phi $,$z)$ for 
the rotating frame $k$. The transformation between the lab and rotating frame 
having angular velocity $\omega $ in the $Z$ direction is

\begin{equation}
\label{eq2}
\begin{array}{l}
 cT = ct \\ 
 R = r \\ 
 \Phi = \phi + \omega t \\ 
 Z = z\;. \\ 
 \end{array}
\end{equation}

In matrix form this may be expressed as

\begin{equation}
\label{eq3}
\Lambda ^\alpha _B = \left[ {{\begin{array}{*{20}c}
 1 \hfill & 0 \hfill & 0 \hfill & 0 \hfill \\
 0 \hfill & 1 \hfill & 0 \hfill & 0 \hfill \\
 { - \textstyle{\omega \over c}} \hfill & 0 \hfill & 1 \hfill & 0 \hfill \\
 0 \hfill & 0 \hfill & 0 \hfill & 1 \hfill \\
\end{array} }} \right]\quad \quad \quad \quad \quad \Lambda ^A_\beta = 
\left[ {{\begin{array}{*{20}c}
 1 \hfill & 0 \hfill & 0 \hfill & 0 \hfill \\
 0 \hfill & 1 \hfill & 0 \hfill & 0 \hfill \\
 {\textstyle{\omega \over c}} \hfill & 0 \hfill & 1 \hfill & 0 \hfill \\
 0 \hfill & 0 \hfill & 0 \hfill & 1 \hfill \\
\end{array} }} \right],
\end{equation}

\noindent
where $A$ and $B$ here are upper case Greek, $\Lambda ^{\alpha }_{B}$ 
transforms a contravariant lab vector [e.g., \textit{dX}$^{B}$ = (\textit{cdT,dR,d}$\Phi $\textit{,dZ})$^{T}$] to 
the corresponding contravariant rotating frame vector [\textit{dx}$^{\alpha }$ = 
(\textit{cdt,dr,d}$\phi $\textit{,dz})$^{T}$], and $\Lambda ^{A}_{\beta }$ transforms the latter back 
from the rotating frame to the lab.

The following relations, which we will use in subsequent sections, were 
first championed by Langevin\cite{Paul:1937}, can be found in many 
sources\cite{Ref:5}, and are shown in Klauber\cite{Ref:6} to be derivable 
from (\ref{eq2}). The rotating frame coordinate metric $g_{\alpha \beta }$ and its 
inverse $g^{\alpha \beta }$ are

\begin{equation}
\label{eq4}
g_{\alpha \beta } = \left[ {{\begin{array}{*{20}c}
 { - (1 - \textstyle{{r^2\omega ^2} \over {c^2}})} \hfill & 0 \hfill & 
{\textstyle{{r^2\omega } \over c}} \hfill & 0 \hfill \\
 0 \hfill & 1 \hfill & 0 \hfill & 0 \hfill \\
 {\textstyle{{r^2\omega } \over c}} \hfill & 0 \hfill & {r^2} \hfill & 0 
\hfill \\
 0 \hfill & 0 \hfill & 0 \hfill & 1 \hfill \\
\end{array} }} \right]\quad \quad \quad \quad \quad \quad g^{\alpha \beta } 
= \left[ {{\begin{array}{*{20}c}
 { - 1} \hfill & 0 \hfill & {\textstyle{\omega \over c}} \hfill & 0 \hfill 
\\
 0 \hfill & 1 \hfill & 0 \hfill & 0 \hfill \\
 {\textstyle{\omega \over c}} \hfill & 0 \hfill & {\textstyle{{(1\;\; - 
\;\;\textstyle{{r^2\omega ^2} \over {c^2}})} \over {r^2}}} \hfill & 0 \hfill 
\\
 0 \hfill & 0 \hfill & 0 \hfill & 1 \hfill \\
\end{array} }} \right].
\end{equation}

The off-diagonal terms imply that time and space are not orthogonal in the 
rotating frame. Although a little unusual, the rotating frame metric is not 
alone in this regard, and shares this NTO characteristic with the spacetime 
metric around a massive body such as a star or black hole that possesses 
angular momentum\cite{Charles:1973}.

For completeness, the lab metric $G_{AB}$ and its inverse $G^{AB}$ are

\begin{equation}
\label{eq5}
G_{AB} = \left[ {{\begin{array}{*{20}c}
 { - 1} \hfill & 0 \hfill & 0 \hfill & 0 \hfill \\
 0 \hfill & 1 \hfill & 0 \hfill & 0 \hfill \\
 0 \hfill & 0 \hfill & {R^2} \hfill & 0 \hfill \\
 0 \hfill & 0 \hfill & 0 \hfill & 1 \hfill \\
\end{array} }} \right]\quad \quad \quad \quad \quad G^{AB} = = \left[ 
{{\begin{array}{*{20}c}
 { - 1} \hfill & 0 \hfill & 0 \hfill & 0 \hfill \\
 0 \hfill & 1 \hfill & 0 \hfill & 0 \hfill \\
 0 \hfill & 0 \hfill & {\textstyle{1 \over {R^2}}} \hfill & 0 \hfill \\
 0 \hfill & 0 \hfill & 0 \hfill & 1 \hfill \\
\end{array} }} \right].
\end{equation}

\subsection{Mathematical vs. measured components}
\label{subsec:mathematical}

When working with rotating frames, we need to keep two things in mind that 
are usually irrelevant for Minkowski metrics in Lorentz frames, but are 
quite relevant for NTO frames. Both of these concern the relationship 
between generalized components of four-vectors (i.e., the mathematical 
components one works with in analyses, which are called \textit{coordinate} components) and\textit{ physical} 
components (i.e., the components one would actually measure with standard 
instruments in an experiment.) Contravariant/covariant coordinate components 
of a vector do not equal physical components except for the special case 
where the coordinate system basis is orthonormal, such as in Minkowski 
coordinates.

\subsection{Contravariant vs. covariant four-vectors}

The first of the aforementioned concerns lies with the covariant or 
contravariant nature of the vector components.\cite{Robert:2001} 
Generalized coordinates (e.g., $x^{\alpha })$ are expressed as contravariant 
quantities, and generalized four-velocity $u^{\alpha }$ is simply the 
derivative of these coordinates with respect to the invariant scalar 
quantity $\tau _{p} $ (proper time of the particle.) In the 
strictest and most general sense, four-velocities only represent (proper) 
time derivatives of the coordinates if they are expressed in contravariant 
form. For example, in an NTO frame lowering the index of $u^{\alpha }$ via 
the metric $g_{\alpha \beta }$ gives components $u_{\alpha }$ which are \textit{not} the 
proper time derivatives of their respective coordinate values. This is true 
because $g_{\alpha \beta }$ is not the identity matrix. Note that in 
Minkowski coordinates $g_{\alpha \beta }=\eta _{\alpha \beta }$, which 
is, apart from the sign of the $g_{00}$ component, an identity matrix. In a 
coordinate frame with such a Minkowski metric the covariant form of the 
four-velocity is identical to the contravariant form except for the sign of 
the timelike component. In NTO frames, however, the difference is much more 
significant, care must be taken, and one must recognize that four-velocity 
is contravariant, not covariant, in form.

Four-momentum, on the other hand, must be treated as a covariant vector. 
This is because the four-momentum is the canonical conjugate of the 
four-velocity. In brief, if the Lagrangian of a given system is

\begin{equation}
\label{eq6}
L\;\; = \;\;L(x^\alpha ,u^\alpha ,\tau _p ),
\end{equation}

\noindent
then the conjugate momentum is

\begin{equation}
\label{eq7}
p_\alpha \; = \;\frac{\partial L}{\partial u^\alpha }.
\end{equation}

This is covariant, not contravariant in form. Hence, it is imperative in an 
NTO system such as a rotating frame that one use covariant components for 
the four-momentum. Contravariant components in such a system, unlike that of 
a system with Minkowski metric, will not represent the physical quantities 
of energy and linear momentum. This is demonstrated explicitly in reference 
\cite{Robert:1998}, section 4.3.4, where it is shown that $p_{0}$, and 
not $p^{0}$, represents the energy of a particle fixed to a rotating disk.

\subsection{Relation between physical and coordinate components}

Getting the correct contravariant or covariant components is not quite 
enough, however, in order to compare theoretical results with measured 
quantities. If a given basis vector does not have unit length, the magnitude 
of the corresponding component will not equal the physical quantity 
measured. For example, a vector with a single non-zero component value of 1 
in a coordinate system where the corresponding basis vector for that 
component has length 3 does not have an absolute (physical) length equal to 
1, but to three.

In general, physical components (those measured with physical instruments in 
the real world) are the components associated with unit basis vectors, and 
generalized coordinate basis vectors are generally not of unit length. As 
shown the Appendix A (see also, texts cited in footnote \cite{The:1972}, 
Malvern\cite{Malvern:1969}, Misner, Thorne and Wheeler\cite{Ref:7}, and 
Klauber\cite{Robert:2002}) physical components are found from generalized 
coordinate components (those used in generalized coordinate mathematical 
analysis) via the relations

\begin{equation}
\label{eq8}
\begin{array}{l}
 v^{\hat {i}{\kern 1pt} } = \sqrt {g_{\underline{i}\underline{i}} } v^i\quad 
\quad v^{\hat {0}{\kern 1pt} } = \sqrt { - g_{00} } v^0 \\ 
 v_{\hat {i}{\kern 1pt} } = \sqrt {g^{\underline{i}\underline{i}}} v_i \quad 
\quad v_{\hat {0}{\kern 1pt} } = \sqrt { - g^{00}} v_0 \\ 
 \end{array},
\end{equation}

\noindent
where carets over indices designate physical quantities, underlining implies 
no summation, Roman indices have values 1,2,3 and the negative signs arise 
on the RHS because $g_{00}$ and $g^{00}$ are negative.

\subsection{Steps in general analysis}
\label{subsec:steps}

Hence, in order to compare theoretical component values with experiment, it 
is necessary to use contravariant components for coordinate differences and 
four-velocity, covariant components for four-momentum, and physical 
components of all component quantities whether covariant or contravariant.

It is important to note, however, that while coordinate components transform 
as true vectors, physical components do not\cite{Fung:1965}. So, while 
physical components are needed to compare theory with experiment, coordinate 
components are needed to carry out vector/tensor analysis.

Steps in NTO analysis therefore comprise

\noindent
i) conversion of known (measured) physical components to coordinate 
components via (\ref{eq8}),

\noindent
ii) appropriate vector/tensor analysis using coordinate components, and

\noindent
iii) conversion of the coordinate component answer back to physical 
component form via (\ref{eq8}) in order to compare with experiment.

We note that the speed of light in (\ref{eq1}) can be derived\cite{Ref:8} using the 
above steps and that said speed is a physical, not coordinate, value.

\section{WAVES IN ROTATING FRAMES}
\label{sec:mylabel2}

\subsection{Overview of procedure}

In order to transform frequencies and wavelengths of light from one frame to 
another we first express those frequencies and wavelengths, via the Planck 
energy and DeBroglie wave relations, as energy and momentum, respectively. 
We use those energy and momentum values to determine appropriate components 
of the generalized four-momentum $p_{\mu }$. We can then simply apply the 
transformations (\ref{eq3}) to transform the four-momentum from the lab to rotating 
frame, and vice versa. Converting the resulting four-momentum components 
back to frequency and wavelength form then reveals Doppler and other wave 
effects from rotation.

\subsection{Lab emission, rotating observer}
\label{subsec:mylabel1}

Consider a photon emitted in the lab in the negative $\Phi $ direction while 
the rotating frame observer is moving in the positive $\Phi $ direction such 
that the observer is approaching the light source. The light with wavelength 
$\lambda $ and frequency $f$ as measured in the lab frame $K$ has four-momentum 
physical components of

\begin{equation}
\label{eq9}
P_{\hat {A}{\kern 1pt} } = \left[ {{\begin{array}{*{20}c}
 { - \frac{hf}{c}} \hfill \\
 0 \hfill \\
 { - \frac{h}{\lambda }} \hfill \\
 0 \hfill \\
\end{array} }} \right],
\end{equation}

\noindent
where $h$ is Planck's constant. From (\ref{eq5}) and (\ref{eq8}), the coordinate components are

\begin{equation}
\label{eq10}
P_A = \left[ {{\begin{array}{*{20}c}
 { - \frac{hf}{c}\frac{1}{\sqrt { - G^{00}} }} \hfill \\
 0 \hfill \\
 { - \frac{h}{\lambda }\frac{1}{\sqrt {G^{22}} }} \hfill \\
 0 \hfill \\
\end{array} }} \right] = \left[ {{\begin{array}{*{20}c}
 { - \frac{hf}{c}} \hfill \\
 0 \hfill \\
 { - \frac{hR}{\lambda }} \hfill \\
 0 \hfill \\
\end{array} }} \right].
\end{equation}

We need to raise the index in order to transform to the rotating frame $k$, as 
our transformations (\ref{eq3}) are specifically for contravariant vectors. With 
(\ref{eq5}), we have

\begin{equation}
\label{eq11}
P^A = G^{AB}P_B = \left[ {{\begin{array}{*{20}c}
 { - 1} \hfill & 0 \hfill & 0 \hfill & 0 \hfill \\
 0 \hfill & 1 \hfill & 0 \hfill & 0 \hfill \\
 0 \hfill & 0 \hfill & {\textstyle{1 \over {R^2}}} \hfill & 0 \hfill \\
 0 \hfill & 0 \hfill & 0 \hfill & 1 \hfill \\
\end{array} }} \right]\left[ {{\begin{array}{*{20}c}
 { - \frac{hf}{c}} \hfill \\
 0 \hfill \\
 { - \frac{hR}{\lambda }} \hfill \\
 0 \hfill \\
\end{array} }} \right] = \left[ {{\begin{array}{*{20}c}
 {\frac{hf}{c}} \hfill \\
 0 \hfill \\
 { - \frac{h}{\lambda R}} \hfill \\
 0 \hfill \\
\end{array} }} \right].
\end{equation}

Using (\ref{eq3}) and $R=r$ from (\ref{eq2}) to transform to the rotating frame, we get

\begin{equation}
\label{eq12}
p^\alpha = \Lambda ^\alpha _B P^B = \left[ {{\begin{array}{*{20}c}
 1 \hfill & 0 \hfill & 0 \hfill & 0 \hfill \\
 0 \hfill & 1 \hfill & 0 \hfill & 0 \hfill \\
 { - \textstyle{\omega \over c}} \hfill & 0 \hfill & 1 \hfill & 0 \hfill \\
 0 \hfill & 0 \hfill & 0 \hfill & 1 \hfill \\
\end{array} }} \right]\left[ {{\begin{array}{*{20}c}
 {\frac{hf}{c}} \hfill \\
 0 \hfill \\
 { - \frac{h}{\lambda R}} \hfill \\
 0 \hfill \\
\end{array} }} \right] = \left[ {{\begin{array}{*{20}c}
 {\frac{hf}{c}} \hfill \\
 0 \hfill \\
 { - \frac{h}{\lambda }\left( {\frac{\omega }{c} + \frac{1}{r}} \right)} 
\hfill \\
 0 \hfill \\
\end{array} }} \right].
\end{equation}

Lowering this to get the necessary covariant form for the four-momentum 
yields

\begin{equation}
\label{eq13}
p_\alpha = g_{\alpha \beta } p^\beta = \left[ {{\begin{array}{*{20}c}
 { - (1 - \textstyle{{r^2\omega ^2} \over {c^2}})} \hfill & 0 \hfill & 
{\textstyle{{r^2\omega } \over c}} \hfill & 0 \hfill \\
 0 \hfill & 1 \hfill & 0 \hfill & 0 \hfill \\
 {\textstyle{{r^2\omega } \over c}} \hfill & 0 \hfill & {r^2} \hfill & 0 
\hfill \\
 0 \hfill & 0 \hfill & 0 \hfill & 1 \hfill \\
\end{array} }} \right]\left[ {{\begin{array}{*{20}c}
 {\frac{hf}{c}} \hfill \\
 0 \hfill \\
 { - \frac{h}{\lambda }\left( {\frac{\omega }{c} + \frac{1}{r}} \right)} 
\hfill \\
 0 \hfill \\
\end{array} }} \right] = \left[ {{\begin{array}{*{20}c}
 { - \frac{hf}{c}\left( {1 + \frac{v}{c}} \right)} \hfill \\
 0 \hfill \\
 { - \frac{hr}{\lambda }} \hfill \\
 0 \hfill \\
\end{array} }} \right].
\end{equation}

We take physical components of (\ref{eq13}) to obtain what an observer in the 
rotating frame would measure with physical instruments, i.e.,

\begin{equation}
\label{eq14}
p_{\hat {\alpha }{\kern 1pt} } = \left[ {{\begin{array}{*{20}c}
 { - \sqrt { - g^{00}} \frac{hf}{c}\left( {1 + \frac{v}{c}} \right)} \hfill 
\\
 0 \hfill \\
 { - \sqrt {g^{22}} \frac{hr}{\lambda }} \hfill \\
 0 \hfill \\
\end{array} }} \right] = \left[ {{\begin{array}{*{20}c}
 { - \frac{hf}{c}\left( {1 + \frac{v}{c}} \right)} \hfill \\
 0 \hfill \\
 { - \frac{h\sqrt {1 - v^2 / c^2} }{\lambda }} \hfill \\
 0 \hfill \\
\end{array} }} \right].
\end{equation}

We then need to express (\ref{eq14}) in terms of the frequency $f_{k}$ and wavelength 
$\lambda _{k}$ measured by an observer in the rotating frame k. We can not 
simply assume the Planck and DeBrolie relations hold in a rotating frame, as 
there is no guarantee that they have the same form in an NTO frame. That is, 
we can not presume that $p_{\hat {0}{\kern 1pt} } c = E_k $ equals $hf_k $, 
nor that $p_{\hat {2}{\kern 1pt} } $ equals $h / \lambda _k $.

Instead we employ a thought experiment and physical world logic. We know 
time in the rotating frame runs more slowly than time in the lab by the 
inverse of the Lorentz factor $1 / \gamma = \sqrt {1 - v^2 / c^2} $, and 
this should increase the frequency of light as seen on the rotating frame 
($f_{k})$ over that of the lab ($f)$. Independent of that, a rotating frame 
observer moving toward a lab oscillator should see that oscillator beating 
faster than would an observer in the lab. At least at low speeds, this 
increase should be by the factor $(1 + v / c) = (1 + \omega r / c)$, as 
shown in Tables 1 and 2. Based on these physical considerations, we propose 
a relationship between lab and rotating frame frequencies for a given photon 
of light as

\begin{equation}
\label{eq15}
f_k = \frac{f\left( {1 + \frac{v}{c}} \right)}{\sqrt {1 - v^2 / c^2} }.
\end{equation}

Given (\ref{eq15}) and the fact that $f_k \lambda _k = u_{light,circum} $ of (\ref{eq1}), it 
follows that

\begin{equation}
\label{eq16}
\lambda _k = \lambda .
\end{equation}

Thus (\ref{eq14}) becomes, in terms of rotating frame observable quantities (RHS 
below),

\begin{equation}
\label{eq17}
p_{\hat {\alpha }{\kern 1pt} } = \left[ {{\begin{array}{*{20}c}
 { - \frac{hf}{c}\left( {1 + \frac{v}{c}} \right)} \hfill \\
 0 \hfill \\
 { - \frac{h\sqrt {1 - v^2 / c^2} }{\lambda }} \hfill \\
 0 \hfill \\
\end{array} }} \right] = \left[ {{\begin{array}{*{20}c}
 { - \frac{hf_k \sqrt {1 - v^2 / c^2} }{c}} \hfill \\
 0 \hfill \\
 { - \frac{h\sqrt {1 - v^2 / c^2} }{\lambda _k }} \hfill \\
 0 \hfill \\
\end{array} }} \right].
\end{equation}

From this the Planck and DeBroglie relations for the NTO rotating disk frame 
are seen to be

\begin{equation}
\label{eq18}
E_k = hf_k \sqrt {1 - v^2 / c^2} \quad \quad \quad p_{phys,circum} = 
\frac{h\sqrt {1 - v^2 / c^2} }{\lambda _k }.
\end{equation}

Similar results for the radially directed light are derived in Appendix B.

The frequency and wavelength results above are summarized in Case 1 of Table 
3. In Table 3, non-subscripted quantities and velocity $V$ refer to the lab 
frame $K$, and the subscript $k$ designates rotating frame quantities. Note if $v$ 
were in the negative $\Phi $ direction, or if the rotating frame observer 
were moving away from the light emission source in the lab, then $v \to - v$ 
in (\ref{eq15}) and throughout Table 3.

\subsection{Rotating frame emission, lab observer}
\label{subsec:rotating}

Consider now a photon emitted from the rotating frame in the positive $\Phi 
$ direction from a source on the rotating frame that is approaching a lab 
observer. We can simply reverse the steps (\ref{eq14}) to (\ref{eq9}) of Section 
\ref{subsec:mylabel1} to relate wavelengths and frequencies in the 
rotating and lab frames, taking care that the sign for linear momentum 
changes from the earlier case. The reader can either carry out these steps 
to justify the results summarized in Case 2 of Table 3, or consider the 
following logic.

We know from (\ref{eq1}) that the speed of light in the rotating frame in this case 
is

\begin{equation}
\label{eq19}
u_{light,circum} = v_k = \frac{c - v}{\sqrt {1 - v^2 / c^2} }.
\end{equation}

We also know that frequency should increase by first order in $v/c$ as seen by 
the lab observer because the source is approaching, yet it should decrease 
by the inverse Lorentz factor since time runs more slowly in the rotating 
frame. Further, the wavelengths in the rotating and lab frames should be 
related mathematically in the very same way as (\ref{eq16}). Thus,

\begin{equation}
\label{eq20}
f_k = \frac{f(1 - v / c)}{\sqrt {1 - v^2 / c^2} }\quad \quad \quad \lambda 
_k = \lambda ,
\end{equation}

\noindent
such that (\ref{eq19}) holds, i.e.,

\begin{equation}
\label{eq21}
f_k \lambda _k = \frac{f\lambda (1 - v / c)}{\sqrt {1 - v^2 / c^2} } = 
\frac{c - v}{\sqrt {1 - v^2 / c^2} } = v_k .
\end{equation}

Hence, frequency seen in the lab from emission in the rotating frame is

\begin{equation}
\label{eq22}
f = \frac{f_k \sqrt {1 - v^2 / c^2} }{(1 - v / c)} = f_k (1 + v / c - 
\textstyle{1 \over 2}v^2 / c^2 + ...).
\end{equation}

These results are summarized in the right hand column of Table 3.

The relation (\ref{eq22}) could also have been deduced directly from the LHS of (\ref{eq15}) 
with the realization that changing the direction of the photon has the same 
effect mathematically as changing the direction of rotation. That is $v \to 
- v$ in (\ref{eq15}) yields (\ref{eq22}). Note that higher order Doppler shift effects for 
rotating frames differ from those of both Sections 
\ref{sec:newtonian} and \ref{sec:waves}.

\begin{center}
\textbf{Table 3. Circumferentially Directed Waves in Relativistically 
Rotating Frames}
\end{center}

\begin{longtable}
{|p{85pt}|p{166pt}|p{157pt}|}
a & a & a  \kill
\hline
& 
\textbf{Case 1}& 
\textbf{Case 2} \\
\hline
Source& 
Lab frame K& 
Rotating frame k \\
\hline
Observer& 
Rotating frame k& 
Lab frame K \\
\hline
Motion& 
Observer toward \par source at $v=\omega r$& 
Source toward \par observer at $v=\omega r$ \\
\hline
Wave speed& 
$V = c_{\kern 1pt} ^{\kern 1pt} $ \par $v_k = \frac{c + v}{\sqrt {1 - v^2 / c^2} }$& 
$V = c_{\kern 1pt} ^{\kern 1pt} $ \par $v_k = \frac{c - v}{\sqrt {1 - v^2 / c^2} }$ \\
\hline
Wave length& 
$\lambda _k = \lambda $& 
$\lambda _k = \lambda $ \\
\hline
Frequency& 
$f_k = \frac{f\left( {1 + v / c} \right)}{\sqrt {1 - v^2 / c^2} }$& 
$f = \frac{f_k \sqrt {1 - v^2 / c^2} }{1 - v / c}$ \\
\hline
Doppler shift \par observer sees& 
$f_k = f\left( {1 + v / c - \textstyle{1 \over 2}v^2 / c^2 + ...} \right)_{\kern 1pt} ^{\kern 1pt} $& 
$f = f_k \left( {1 + v / c + \textstyle{1 \over 2}v^2 / c^2 + ...} \right)_{\kern 1pt} ^{\kern 1pt} $ \\
\hline
\label{tab3}
\end{longtable}

\section{TURNING THE LIGHT SOURCE IN THE ROTATING FRAME}
\label{sec:turning}

For completeness, we consider the effect on light speed, frequency and 
wavelength measurements in the rotating frame when a source in the rotating 
frame is turned. This can be valuable for evaluation of Brillet and 
Hall\cite{Brillet:1979} type experiments in which test apparatus is turned 
within the rotating frame of the earth. One part of the apparatus is the 
light source; another part the sensing equipment.

We consider two cases: the circumferential direction and the radial 
direction. The vertical ($z)$ direction parallels that of the radial direction. 
We take it as an axiom that an observer who is fixed relative to a source 
detects no Doppler shift in frequency from that source. This axiom appears 
to have been tested to extremely high accuracy by Chen and 
Liu\cite{Shaoguang:1} although there are certain caveats\cite{The:1} 
regarding the relevance of that test to the present article.

\subsection{Circumferential Direction}

In the circumferential direction, light speed is given by (\ref{eq1}). For observers 
on the rotating frame the frequency of light emitted from a source on the 
rotating frame is $f_{k}$. Hence, the wavelength in the circumferential 
direction, determined solely by measurements on the rotating frame must be

\begin{equation}
\label{eq23}
\lambda _{k,circum} = \frac{u_{light,circum} }{f_k } = \frac{c\pm v}{f_k 
\sqrt {1 - v^2 / c^2} }.
\end{equation}

\subsection{Radial Direction}
\label{subsec:radial}

\subsubsection{Radial Direction Light Speed}

From the rotating (NTO) frame metric of (\ref{eq4}), the line element is

\begin{equation}
\label{eq24}
\begin{array}{c}
 ds^2 = - c^2(1 - r^2\omega ^2 / c^2)dt^2 + dr^2 + r^2d\phi ^2 + 2r^2\omega 
d\phi dt + dz^2 \\ 
 \quad \\ 
 = g_{\alpha \beta } dx^\alpha dx^\beta \\ 
 \end{array}
\end{equation}

\noindent
where $t$ is coordinate time and equals that on standard clocks in the lab. 
Note that the time on a standard clock at a fixed 3D location on the 
rotating disk, found by taking \textit{ds}$^{2}= -c^{2}d\tau $ and \textit{dr = d}$\phi $ = \textit{dz = }0, is

\begin{equation}
\label{eq25}
d\tau = d\hat {t} = \sqrt {1 - r^2\omega ^2 / c^2} dt,
\end{equation}

\noindent
where the caret over \textit{dt} indicates \textit{physical} time (i.e., time measured with standard 
clocks fixed in the rotating frame.)

For a radially directed ray of light, \textit{d$\phi $} = \textit{dz = }0, and \textit{ds} = 0. Solving for \textit{dr/dt} one 
obtains

\begin{equation}
\label{eq26}
\frac{dr}{dt} = c\sqrt {1 - r^2\omega ^2 / c^2} .
\end{equation}

Since $g_{rr}$ = 1, the physical component, (measured with standard meter 
sticks) for radial displacement $d\hat {r}$ equals the coordinate radial 
displacement \textit{dr}. The physical (measured) speed of light in the radial 
direction is therefore

\begin{equation}
\label{eq27}
v_{light,radial,phys} = \frac{d\hat {r}}{d\hat {t}} = \frac{dr}{d\tau } = 
\frac{dr}{\sqrt {1 - r^2\omega ^2 / c^2} dt} = c.
\end{equation}

This can be further justified by physical considerations of the path 
followed by such a light ray as seen in the lab frame, along with (\ref{eq25}).

\subsubsection{Radial Direction Wavelength}
\label{subsubsec:radial}

As the apparatus containing the light source and detector is turned from the 
circumferential to radial direction, the speed of light in the rotating 
frame changes from that of (\ref{eq1}) to that of (\ref{eq27}). The frequency remains 
unchanged. Hence, the wavelength changes to

\begin{equation}
\label{eq28}
\lambda _{k,radial} = \frac{c}{f_k }.
\end{equation}

From (\ref{eq23}) we see that

\begin{equation}
\label{eq29}
\lambda _{k,circum} = \frac{1\pm v / c}{\sqrt {1 - v^2 / c^2} }\lambda 
_{k,radial} .
\end{equation}

The results of this and the prior section are summarized in Table 4.

\bigskip

\begin{center}
\textbf{Table 4. Turning Light Source in Relativistically Rotating Frame}
\end{center}

\bigskip

\begin{longtable}[htbp]
{|p{76pt}|p{198pt}|p{193pt}|}
a & a & a  \kill
\hline
 & 
\textbf{Case 1} \par \textbf{Circumferential Direction}& 
\textbf{Case 2} \par \textbf{Radial Direction} \\
\hline
Source& 
Rotating frame k& 
Rotating frame k \\
\hline
Observer& 
Rotating frame k& 
Rotating frame k \\
\hline
Motion& 
Source and observer \par both fixed in k & 
Source and observer \par both fixed in k  \\
\hline
Wave speed& 
$v_{k,circum} = \frac{c\pm v}{\sqrt {1 - v^2 / c^2} }_{\kern 1pt} ^{\kern 1pt} $& 
$v_{k,radial} = c$ \\
\hline
Wave length& 
$\lambda _{observ,circum} = \lambda _{source,circum} = \,\lambda _{k,circum} $ \par $\lambda _{k,circum} = \frac{c\pm v}{f_k \sqrt {1 - v^2 / c^2} } = \frac{1\pm v / c}{\sqrt {1 - v^2 / c^2} }\lambda _{k,radial} $& 
$\lambda _{observ,radial} = \lambda _{source,radial} = \,\lambda _{k,radial} $ \par $\lambda _{k,radial} = \frac{c}{f_k }$ \\
\hline
Frequency& 
$f_{observ,circum} = f_{source,circum} = f_{k,circum} = f_k $& 
$f_{observ,radial} = f_{source,radial} = f_{k,radial} = f_k $ \\
\hline
Doppler shift \par observer sees& 
None& 
None \\
\hline
\label{tab4}
\end{longtable}

\subsection{Possible First Order Test}
\label{subsec:possible}

The difference in wavelength between the radial and circumferential 
directions shown above is first order in $v/c$. Although tests of light speed 
\textit{per se} are limited to round trip measurements, and hence must be of second order, 
there are a number of tests, such as the Young double slit experiment, that 
detect differences in wavelength. As readily seen in (\ref{eq23}) and (\ref{eq28}) above, 
wavelength is proportional to one way light speed. Thus, a test that 
measures wavelength could be a first order test of light speed. Such an 
experiment could be performed using the earth as the rotating frame and 
turning the test apparatus relative to the earth.

\section{SUMMARY AND CONCLUSIONS}

Comparison of Tables 1, 2 and 3 shows that to first order rotating frame 
Doppler effects are equivalent to those of classical and relativistic 
analyses. Higher order effects, however, vary among the three. As seen in 
Table 4, turning the light source within the rotating frame changes light 
speed and wavelength, but not frequency, for rotating frame based observers. 
Multiplication of frequency by wavelength in all cases equals wave speed, 
providing corroboration for the methodology employed.
\appendix
\section*{APPENDIX A. PHYSICAL COMPONENTS}

Consider an arbitrary vector \textbf{v} in a 2D space

\begin{equation}
\label{eq30}
{\rm {\bf v}} = v^1{\rm {\bf e}}_1 + v^2{\rm {\bf e}}_2 = v^{\hat {1}{\kern 
1pt} }{\rm {\bf \hat {e}}}_1 + v^{\hat {2}{\kern 1pt} }{\rm {\bf \hat 
{e}}}_2 
\end{equation}

\noindent
where \textbf{e}$_{i}$ are coordinate basis vectors and ${\rm {\bf \hat 
{e}}}_i $ are unit length (non-coordinate) basis vectors pointing in the 
same respective directions. That is,

\begin{equation}
\label{eq31}
{\rm {\bf \hat {e}}}_i = \frac{{\rm {\bf e}}_i }{\vert {\rm {\bf e}}_i \vert 
} = \frac{{\rm {\bf e}}_i }{\sqrt {{\rm {\bf e}}_{\underline{i}} \cdot {\rm 
{\bf e}}_{\underline{i}} } } = \frac{{\rm {\bf e}}_i }{\sqrt 
{g_{\underline{i}\underline{i}} } }.
\end{equation}

\noindent
where underlining implies no summation. Note that \textbf{e}$_{1}$ and 
\textbf{e}$_{2}$ here do not, in general, have to be orthogonal. Note also, 
that physical components are those associated with unit length basis vectors 
and hence are represented by indices with carets in (\ref{eq30}).

Substituting (\ref{eq31}) into (\ref{eq30}), one readily obtains

\begin{equation}
\label{eq32}
v^{\hat {i}{\kern 1pt} } = \sqrt {g_{\underline{i}\underline{i}} } v^i.
\end{equation}

Relation (\ref{eq32}) between physical and coordinate components is valid locally 
in curved, as well as flat, spaces and can be extrapolated to 4D general 
relativistic applications, to higher order tensors, and to covariant 
components.

\section*{APPENDIX B. RADIAL DIRECTION: LAB VS. ROTATING FRAME}
\label{sec:appendix}

We would be remiss if we did not include an analysis similar to that of 
sections \ref{subsec:mylabel1} and \ref{subsec:rotating} for 
the radial direction. We outline this below.

\subsection*{Photon in Lab Frame Radial Direction}

Parallel to (\ref{eq9}) through (\ref{eq18}), we consider a photon emitted in the lab in the 
$R$ direction. We determine the physical components of the 4 momentum vector in 
the lab, convert to coordinate components, raise the index, transform to the 
rotating frame, lower the index, and find physical components.

\begin{equation}
\label{eq33}
\begin{array}{l}
 p_{\hat {A}{\kern 1pt} } = \left[ {{\begin{array}{*{20}c}
 { - \textstyle{{hf} \over c}} \hfill \\
 {\textstyle{h \over \lambda }} \hfill \\
 0 \hfill \\
 0 \hfill \\
\end{array} }} \right]\; \to \;p_A = \left[ {{\begin{array}{*{20}c}
 { - \textstyle{{hf} \over c}} \hfill \\
 {\textstyle{h \over \lambda }} \hfill \\
 0 \hfill \\
 0 \hfill \\
\end{array} }} \right]\; \to \;p^A = \left[ {{\begin{array}{*{20}c}
 { - \textstyle{{hf} \over c}} \hfill \\
 {\textstyle{h \over \lambda }} \hfill \\
 0 \hfill \\
 0 \hfill \\
\end{array} }} \right] \\ 
 \\ 
 \to \;p^\alpha = \left[ {{\begin{array}{*{20}c}
 { - \textstyle{{hf} \over c}} \hfill \\
 {\textstyle{h \over \lambda }} \hfill \\
 { - \textstyle{{\omega hf} \over {c^2}}} \hfill \\
 0 \hfill \\
\end{array} }} \right]\; \to \;p_\alpha = \left[ {{\begin{array}{*{20}c}
 { - \textstyle{{hf} \over c}} \hfill \\
 {\textstyle{h \over \lambda }} \hfill \\
 0 \hfill \\
 0 \hfill \\
\end{array} }} \right]\; \to \;p_{\hat {\alpha }{\kern 1pt} } = \left[ 
{{\begin{array}{*{20}c}
 { - \textstyle{{hf} \over c}} \hfill \\
 {\textstyle{h \over \lambda }} \hfill \\
 0 \hfill \\
 0 \hfill \\
\end{array} }} \right] \\ 
 \end{array}
\end{equation}

We conclude that the speed, energy and three-momentum of a light ray 
directed radially in the lab is identical to that measured on the rotating 
disk. This is interesting since this photon as seen in the rotating frame is 
not moving in a purely radial direction, but has a circumferential 
component. To prove this we can not simply transform the four velocity for 
the photon since four velocity $U^A = dX^A / d\tau _p $, where \textit{$\tau $}$_{p}$ is the 
proper length of the path in spacetime, and for a photon \textit{d$\tau $}$_{p}$ = 0. 
Consider instead the displacement of the photon having three velocity in the 
radial direction $dR / dT = c$. The 4D displacement vector in time \textit{dT} shown on 
the left side of (\ref{eq34}) below can then be transformed to the rotating frame 
and the physical components determined in that frame as follows

\begin{equation}
\label{eq34}
dX^{\hat {A}{\kern 1pt} } = dX^A = \left[ {{\begin{array}{*{20}c}
 {cdT} \hfill \\
 {cdT} \hfill \\
 0 \hfill \\
 0 \hfill \\
\end{array} }} \right]\quad \to \quad dx^\alpha = \Lambda ^\alpha _A dX^A = 
\left[ {{\begin{array}{*{20}c}
 {cdt} \hfill \\
 {cdt} \hfill \\
 { - \omega dt} \hfill \\
 0 \hfill \\
\end{array} }} \right]\quad \to \quad dx^{\hat {\alpha }{\kern 1pt} } = 
\left[ {{\begin{array}{*{20}c}
 {\sqrt {1 - \omega ^2r^2 / c^2} cdt} \hfill \\
 {cdt} \hfill \\
 { - \omega rdt} \hfill \\
 0 \hfill \\
\end{array} }} \right].
\end{equation}

The physical time of a standard clock fixed in the rotating 
frame\cite{See:1} is $d\tau = \sqrt {1 - v^2 / c^2} dt$ where \textit{v=$\omega $r}. Hence the 
physical speeds of the photon in the rotating frame in the radial and 
circumferential directions are, respectively,

\begin{equation}
\label{eq35}
v_{light,phys,radial} = \frac{dx^{\hat {1}{\kern 1pt} }}{d\tau } = 
\frac{cdt}{d\tau } = \frac{c}{\sqrt {1 - v^2 / c^2} },\quad \quad 
v_{light,phys,circum} = \frac{dx^{\hat {2}{\kern 1pt} }}{d\tau } = \frac{ - 
vdt}{d\tau } = \frac{ - v}{\sqrt {1 - v^2 / c^2} }.
\end{equation}

\subsection*{Photon in Rotating Frame Radial Direction}
\label{subsec:photon}

Consider, on the other hand, a photon in the lab with a circumferential 
speed component of $v=\omega r$. One might expect this to travel along a purely 
radial direction in the rotating frame, and it does, as we prove at the end 
of this appendix. Note that the radial velocity component in the lab must be 
$c\sqrt {1 - v^2 / c^2} $ in order for the photon to have speed $c$ along its 
direction of travel in the lab.

The energy and momentum in the rotating frame for this photon may be found 
in parallel fashion to that for the lab purely radial direction photon 
above. Starting with the physical 4 momentum in the lab we find

\begin{equation}
\label{eq36}
\begin{array}{l}
 p_{\hat {A}{\kern 1pt} } = \left[ {{\begin{array}{*{20}c}
 { - \textstyle{{hf} \over c}_{\kern 1pt} ^{\kern 1pt} } \hfill \\
 {\sqrt {1 - v^2 / c^2} \textstyle{h \over \lambda }_{\kern 1pt} ^{\kern 
1pt} } \hfill \\
 {\textstyle{v \over c}\textstyle{h \over \lambda }_{\kern 1pt} ^{\kern 1pt} 
} \hfill \\
 0 \hfill \\
\end{array} }} \right]\; \to \;p_A = \left[ {{\begin{array}{*{20}c}
 { - \textstyle{{hf} \over c}_{\kern 1pt} ^{\kern 1pt} } \hfill \\
 {\sqrt {1 - v^2 / c^2} \textstyle{h \over \lambda }_{\kern 1pt} ^{\kern 
1pt} } \hfill \\
 {R\textstyle{v \over c}\textstyle{h \over \lambda }_{\kern 1pt} ^{\kern 
1pt} } \hfill \\
 0 \hfill \\
\end{array} }} \right]\; \to \;p^A = \left[ {{\begin{array}{*{20}c}
 {\textstyle{{hf} \over c}_{\kern 1pt} ^{\kern 1pt} } \hfill \\
 {\sqrt {1 - v^2 / c^2} \textstyle{h \over \lambda }_{\kern 1pt} ^{\kern 
1pt} } \hfill \\
 {\textstyle{1 \over R}\textstyle{v \over c}\textstyle{h \over \lambda 
}_{\kern 1pt} ^{\kern 1pt} } \hfill \\
 0 \hfill \\
\end{array} }} \right] \\ 
 \\ 
 \to \;p^\alpha = \left[ {{\begin{array}{*{20}c}
 {\textstyle{{hf} \over c}_{\kern 1pt} ^{\kern 1pt} } \hfill \\
 {\sqrt {1 - v^2 / c^2} \textstyle{h \over \lambda }_{\kern 1pt} ^{\kern 
1pt} } \hfill \\
 0 \hfill \\
 0 \hfill \\
\end{array} }} \right] \to \;p_\alpha = \left[ {{\begin{array}{*{20}c}
 { - (1 - v^2 / c^2)\textstyle{{hf} \over c}_{\kern 1pt} ^{\kern 1pt} } 
\hfill \\
 {\sqrt {1 - v^2 / c^2} \textstyle{h \over \lambda }_{\kern 1pt} ^{\kern 
1pt} } \hfill \\
 {\textstyle{{vr} \over c}\textstyle{h \over \lambda }_{\kern 1pt} ^{\kern 
1pt} } \hfill \\
 0 \hfill \\
\end{array} }} \right]\; \to \;p_{\hat {\alpha }{\kern 1pt} } = \left[ 
{{\begin{array}{*{20}c}
 { - (1 - v^2 / c^2)\textstyle{{hf} \over c}_{\kern 1pt} ^{\kern 1pt} } 
\hfill \\
 {\sqrt {1 - v^2 / c^2} \textstyle{h \over \lambda }_{\kern 1pt} ^{\kern 
1pt} } \hfill \\
 {\sqrt {1 - v^2 / c^2} \textstyle{v \over c}\textstyle{h \over \lambda 
}_{\kern 1pt} ^{\kern 1pt} } \hfill \\
 0 \hfill \\
\end{array} }} \right]. \\ 
 \end{array}
\end{equation}

As a check, one can see that $p^{2}$ = 0 in both frames, as it must. The 
differences in physical values of energy and momentum between the lab and 
rotating frames are second order. And though the photon has purely radial 
direction of travel in the rotating frame, it has a circumferential momentum 
component in that frame\cite{The:2}.

We can then, as before, determine the physical 4 displacement in time \textit{dT} in 
the lab, convert to coordinate displacement, transform to the rotating 
frame, and find the physical components in that frame.

\begin{equation}
\label{eq37}
\begin{array}{l}
 dX^{\hat {A}{\kern 1pt} } = \left[ {{\begin{array}{*{20}c}
 {cdT_{\kern 1pt} ^{\kern 1pt} } \hfill \\
 {c\sqrt {1 - v^2 / c^2} dT_{\kern 1pt} ^{\kern 1pt} } \hfill \\
 {vdT_{\kern 1pt} ^{\kern 1pt} } \hfill \\
 0 \hfill \\
\end{array} }} \right]\;\; \to \;\;dX^A = \left[ {{\begin{array}{*{20}c}
 {cdT_{\kern 1pt} ^{\kern 1pt} } \hfill \\
 {c\sqrt {1 - v^2 / c^2} dT_{\kern 1pt} ^{\kern 1pt} } \hfill \\
 {\frac{v}{R}dT_{\kern 1pt} ^{\kern 1pt} } \hfill \\
 0 \hfill \\
\end{array} }} \right] \\ 
 \\ 
 \to \;\;dx^\alpha = \left[ {{\begin{array}{*{20}c}
 {cdt_{\kern 1pt} ^{\kern 1pt} } \hfill \\
 {c\sqrt {1 - v^2 / c^2} dt_{\kern 1pt} ^{\kern 1pt} } \hfill \\
 0 \hfill \\
 0 \hfill \\
\end{array} }} \right]\;\; \to \;\;dx^{\hat {\alpha }{\kern 1pt} } = \left[ 
{{\begin{array}{*{20}c}
 {c\sqrt {1 - v^2 / c^2} dt_{\kern 1pt} ^{\kern 1pt} } \hfill \\
 {c\sqrt {1 - v^2 / c^2} dt_{\kern 1pt} ^{\kern 1pt} } \hfill \\
 0 \hfill \\
 0 \hfill \\
\end{array} }} \right] \\ 
 \end{array}
\end{equation}

Thus, we see that the 3 velocity in the rotating frame has only a component 
in the radial direction

\begin{equation}
\label{eq38}
v_{light,phys,circum} = \frac{c\sqrt {1 - v^2 / c^2} dt}{d\tau } = c,
\end{equation}

\noindent
as asserted, and this agrees with the earlier result (\ref{eq27}).


\begin{thebibliography}{19}
\bibitem{Robert:1998} Robert D. Klauber, ``New perspectives on the relatively rotating disk and non-time-orthogonal reference frames", \textit{Found. Phys. Lett}. 11(\ref{eq5}), 405-443 (Oct 1998), gr-qc/0103076
\bibitem{Robert:1999}Robert D. Klauber, ``Comments regarding recent articles on relativistically rotating frames'', \textit{Am. J. Phys}. 67(\ref{eq2}), 158-159 ( Feb 1999), gr-qc/9812025.
\bibitem{Robert:1}Robert D. Klauber, ``Non-time-orthogonality and tests of special relativity'', gr-qc/0006023. 
\bibitem{Ref:1} Ref \cite{Robert:1998}, Section 4.3, pp. 425-429.
\bibitem{Ref:2} Ref. \cite{Robert:1998}, Section 6, pp. 434-436.
\bibitem{Ref:3}Ref. \cite{Robert:1}, Section 5.
\bibitem{Neil:1997} Neil Ashby, ``Relativistic Effects in the Global Positioning System'', \textit{15}$^{th}$\textit{ Intl. Conf. Gen. Rel. and Gravitation,} Pune, India (Dec 15-21, 1997), available at www.colorado.edu/engineering/GPS/Papers/RelativityinGPS.ps. See pp. 5-7.
\bibitem{Neil:2002}Neil Ashby, ``Relativity and the Global Positioning System'', \textit{Phys. Today}, May 2002, 41-47. See pg 44.
\bibitem{Robert:2}Robert D. Klauber, ``Non-time-orthogonal reference frames in the theory of relativity'', gr-qc/0005121, section II.
\bibitem{Ref:4} Ref. \cite{Robert:1998}, pg. 425, eq. (\ref{eq19}) modified by the time dilation factor discussed in the subsequent paragraphs therein to yield physical velocity, and pg. 430, eq. (33).
\bibitem{David:1974} David Halliday and Robert Resnick, \textit{Fundamentals of Physics} (John Wiley {\&} Sons, New York, 1974), p.334-337. The authors employ different symbols from those of Table 1. For example, they use \textit{v} as the wave velocity within the medium, whereas we use \textit{v}$_{m}$. They also show that the wavelength in Case 2 of Table 1 is shorter than that of Case 1, whereas we emphasize that in Galilean theory a wave has the same length for the observer and the source frames in each case.
\bibitem{Ibid:1} Ibid., pp.660-663.
\bibitem{Herbert:1941} Herbert E. Ives and G. R. Stilwell, "An Experimental Study of the Rate of a Moving Atomic Clock", \textit{J. Opt. Soc. Am. }28(\ref{eq7}), 215-226 (July 1938). "An Experimental Study of the Rate of Moving Atomic Clock II", \textit{J. Opt. Soc. Am. }B 31, 369-374 (May 1941).
\bibitem{Paul:1937} Paul Langevin, ``Relativit\'{e} -- Sur l'experience de Sagnac'',\textit{ Academie des sciences comptes rendus des seances., }Vol. 205, 304-306 (2 Aug 1937.)
\bibitem{Ref:5} Ref \cite{Neil:1997}, for one.
\bibitem{Ref:6} Ref. \cite{Robert:1998}, Section 4.1, pp. 420-422.
\bibitem{Charles:1973} Charles W. Misner, Kip S. Thorne, and John Archibald Wheeler, \textit{Gravitation }(W. H. Freeman and Co., New York, 1973), Chap. 19, p.448-459.
\bibitem{Robert:2001} Robert D. Klauber, "Generalized Tensor Analysis Method and the Wilson and Wilson Experiment", gr-qc/0107035 (2001). See Section 2.1 and Appendix A therein for more detailed presentation of the physical relevance of contravariant and covariant components in non-orthonormal coordinates.
\bibitem{The:1972}The texts listed below are among those that discuss physical vector and tensor components (the values one measures in experiment) and the relationship between them and coordinate components (the mathematical values that depend on the generalized coordinate system being used.) For a vector this relationship is is $u^{\hat {\mu }\,}\,=\sqrt {g_{\underline{\mu }\underline{\mu }} } u^\mu $ where the caret over the index indicates a physical vector component and underlining implies no summation. See I.S. Sokolnikoff, \textit{Tensor Analysis}, (Wiley {\&} Sons, 1951) pp. 8, 122-127, 205; G.E. Hay, \textit{Vector and Tensor Analysis}, (Dover, 1953) pp 184-186; A. J. McConnell, \textit{Application of Tensor Analysis}, (Dover, 1947) pp. 303-311; Carl E. Pearson, \textit{Handbook of Applied Mathematics}, (Van Nostrand Reinhold, 1983 2$^{nd}$ ed.), pp. 214-216; Murry R. Spiegel, \textit{Schaum's Outline of Vector Analysis}, (Schaum) pg. 172; Robert C. Wrede, \textit{Introduction to Vector and Tensor Analysis,} (Dover 1972), pp. 234-235.
\bibitem{Malvern:1969} L. E. Malvern, \textit{Introduction to the Mechanics of a Continuous Medium} (Prentice-Hall, Englewood Cliffs, New Jersey, 1969), Appendix I, Sec. 5, p.606-613.
\bibitem{Ref:7} Ref. \cite{Charles:1973}. Physical components are introduced on pg. 37, and used in many places throughout the text, e.g. pp. 821-822. Note, however, that the authors confine their explanation and use of physical components to the special case of time orthogonal systems. As noted on page 606 in Ref.\cite{Malvern:1969}, the definitions presented in the present article are applicable to any coordinate system, orthogonal or non-orthogonal, and reduce to those of Ref. \cite{Charles:1973} for orthogonal coordinates.
\bibitem{Robert:2002} Robert D. Klauber, ``Physical Components, Coordinate Components, and the Speed of Light'', gr-qc/0105071, (2001). 
\bibitem{Fung:1965} Y. C. Fung , \textit{Foundations of Solid Mechanics} (Prentice-Hall, Inc., Englewood Cliffs, NJ, 1965), p.53.
\bibitem{Ref:8} Ref. \cite{Robert:1998}, Sections 4.2.5 and 5.1, and Ref. \cite{Robert:2}, Section 5.
\bibitem{Brillet:1979} A. Brillet and J. L. Hall, ``Improved laser test of the isotropy of space,'' \textit{Phys. Rev. Lett}., \textbf{42}(\ref{eq9}), 549-552 (1979).
\bibitem{Shaoguang:1} Shaoguang Chen and Baocheng Liu, ``A New Method for Inspect Spatial Isotropy'', Peking University, http://www.pku.edu.cn/academic/xb/96/e96510.html; ``Experimental Test of the Isotropy of Two-way Speed of Light'', Peking University, http://www.pku.edu.cn/academic/xb/97/{\_}97e509.html. The web sites cited contain only abstracts. The full reports may only be published in the Chinese language. The authors note experimental verification of isotropy of light frequency for $\Delta $\textit{f/f} of 1X10$^{ - 18}$.
\bibitem{The:1} The Chen and Liu tests were for two way light transmission, whereas the present article addresses one way transmission from source to observer. It is also not apparent whether the test apparatus rotated relative to the earth fixed reference frame or the earth frame rotation was used to rotate the apparatus relative to the heavens. This is relevant as NTO analysis predicts the effects described herein only for true rotating frames (in which angular velocity and centrifugal acceleration may be sensed) such as the earth fixed frame. NTO analysis predicts no difference from traditional time-orthogonal relativity theory for gravitational orbits (in which neither angular orbital velocity nor centrifugal acceleration may be sensed.) See ref. \cite{Robert:1998}, section 6.
\bibitem{See:1} See ref. \cite{Robert:1998} Sections 3.3.3, 4.3.1, and 5.1.
\bibitem{The:2} The two cases evaluated in Appendix B suggest that momentum in the inertial lab frame has, to first order, a certain absolute quality to it with respect to rotating frames. This has a parallel in non-relativistic mechanics where the angular momentum of a rotating body (integral of (\textit{$\omega $}\textbf{r x r)}\textit{dm}) calculated in the lab is also the angular momentum seen in the rotating frame of the body itself. This implies linear momentum \textit{$\omega $}\textbf{r}\textit{dm} is the same non-relativistically in both frames as well. This agrees with findings of ref. \cite{Robert:1998}, section 4.3.4 for a massive particle fixed to the disk. The analysis herein therefore appears consistent in the low velocity limit with classical dynamics.
\end{thebibliography}
\end{document}